\documentclass[jou]{interact}

\usepackage{dsfont}
\usepackage{xcolor}
\usepackage{tcolorbox}
\usepackage{cancel}

\usepackage[english]{babel}
\usepackage{inputenc}

\usepackage{multicol, amsthm}

\usepackage{parskip}
\usepackage{booktabs,array}

\newcount\rowc

\makeatletter
\def\ttabular{%
\hbox\bgroup
\let\\\cr
\def\rulea{\ifnum\rowc=\@ne \hrule height 1.3pt \fi}
\def\ruleb{
\ifnum\rowc=1\hrule height 1.3pt \else
\ifnum\rowc=6\hrule height \heavyrulewidth 
   \else \hrule height \lightrulewidth\fi\fi}
\valign\bgroup
\global\rowc\@ne
\rulea
\hbox to 10em{\strut \hfill##\hfill}%
\ruleb
&&%
\global\advance\rowc\@ne
\hbox to 10em{\strut\hfill##\hfill}%
\ruleb
\cr}
\def\endttabular{%
\crcr\egroup\egroup}

\usepackage{epstopdf}% To incorporate .eps illustrations using PDFLaTeX, etc.
\usepackage[caption=false]{subfig}% Support for small, `sub' figures and tables

\usepackage[doublespacing]{setspace}% To produce a `double spaced' document if required
\usepackage{inputenc}

\theoremstyle{plain}% Theorem-like structures provided by amsthm.sty

\theoremstyle{definition}

\theoremstyle{remark}

\usepackage{natbib}

\begin{document}

\bibliographystyle{apalike}

\articletype{Original Research Paper}% Specify the article type or omit as appropriate

\title{\emph{Can we disregard the whole model?} Omnibus non-inferiority testing for $R^{2}$ in multivariable linear regression and $\hat{\eta}^{2}$ in ANOVA.}

\author{
\name{Harlan Campbell\textsuperscript{a}\thanks{CONTACT Harlan Campbell. Email: harlan.campbell@stat.ubc.ca}, Dani\"{e}l Lakens\textsuperscript{b}}
\affil{\textsuperscript{a}University of British Columbia, Department of Statistics}
\affil{\textsuperscript{b}Eindhoven University of Technology}
}

\maketitle

\begin{abstract}
Determining a lack of association between an outcome variable and a number of different explanatory variables is frequently necessary in order to disregard a proposed model (i.e., to confirm the lack of a meaningful association between an outcome and predictors). Despite this, the literature rarely offers information about, or technical recommendations concerning, the appropriate statistical methodology to be used to accomplish this task. This paper introduces non-inferiority tests for ANOVA and linear regression analyses, that correspond to the standard widely used $F$-test for $\hat{\eta}^2$ and $R^{2}$, respectively. A simulation study is conducted to examine the type I error rates and statistical power of the tests, and a comparison is made with an alternative Bayesian testing approach. The results indicate that the proposed non-inferiority test is a potentially useful tool for ``testing the null.'' 
\end{abstract}

\begin{footnotesize}
\noindent The data that support the findings of this study are openly available in the OSF repository ``Can we disregard the whole model?''. at http://doi.org/10.17605/OSF.IO/3Q2VH, reference number 3Q2VH.
\end{footnotesize}

\newpage

\section{Introduction}

 All too often, researchers will conclude that the effect of an explanatory variable, $X$, on an outcome variable, $Y$, is absent when a null-hypothesis significance test (NHST) yields a non-significant $p$-value (e.g., when the $p$-value $> 0.05$).  Unfortunately, such an argument is logically flawed.  As the saying goes, ``absence of evidence is not evidence of absence'' \citep{hartung1983absence, altman1995statistics}.  Indeed, a non-significant result can simply be due to insufficient power, and while a null-hypothesis significance test can provide evidence to \emph{reject} the null hypothesis, it cannot provide evidence \emph{in favour} of the null \citep{quertemont2011statistically}. 
 
 Let $\theta$ be the parameter of interest representing the true association between $X$ and $Y$ in the population of interest.  The equivalence/non-inferiority test reverses the question that is asked in a NHST.  Instead of asking whether we can reject the null hypothesis, e.g., $H_{0}: \theta = 0$, an equivalence test examines whether the magnitude of $\theta$ is at all meaningful: \emph{Can we reject an association between $X$ and $Y$ as large or larger than our smallest effect size of interest, $\Delta$?}  The null hypothesis for an equivalence test is therefore defined as $H_{0}: |\theta| \ge \Delta$.  Or for the one-sided non-inferiority test, the null hypothesis is $H_{0}: \theta \ge \Delta$.  %The alternative hypothesis for the equivalence test is defined, $H_{1}: - \Delta < \theta < \Delta$. 
Note that researchers must decide which effect size is considered meaningful or relevant \citep{lakens2018equivalence}, and define $\Delta$ accordingly, prior to observing any data; see \cite{campbell2018make} for details. 

% is similar to other proposals such as the “three-way testing” scheme proposed by Goeman et al. (2010) [57], and Zhao (2016) [58]’s proposal for incorporating both statistical and clinical significance into one’s testing. 

In a standard multivariable linear regression model, or a standard ANOVA analysis, the variability of the outcome variable, $Y$, is attributed to multiple different explanatory variables, $X_{1}, X_{2}, ..., X_{p}$. Researchers will typically report the linear regression model's $R^{2}$ statistic, or the $\hat{\eta}^{2}$ in the ANOVA context, to estimate the proportion of variance in the observed data that is explained by the model.  \textcolor{black}{To determine whether or not we can reject the hypothesis that the variance attributed to the explanatory variables is equal to zero, one typically calculates an $F$-statistic and tests whether the ``null model'' (i.e., the intercept only model) can be rejected in favour of the ``full model'' (i.e., the model with all explanatory variables included).}  However, in this multivariate setting, while rejecting the ``null model'' is rather simple, concluding \emph{in favour} of the ``null model'' is less obvious.  

If the explanatory variables are not statistically significant, can we simply disregard the full model?  We certainly shouldn't pick and choose which variables to include in the model based on their significance (it is well known that due to model selection bias, most step-wise variable selection schemes are to be avoided; see \cite{hurvich1990impact}). How can we formally test whether the proportion of variance attributable to the full set of explanatory variables is too small to be considered meaningful? In this article, we introduce a non-inferiority test to reject effect sizes that are as large or larger than the smallest effect size of interest as estimated by either the $R^{2}$ statistic or the $\hat{\eta}^{2}$ statistic.

In Section \ref{sec:R2}, we introduce a non-inferiority test for the coefficient of determination parameter in a linear regression context. We show how to define hypotheses and calculate a valid $p$-value for this test based on the $R^{2}$ statistic. We then consider how this frequentist test compares to a Bayesian testing scheme based on Bayes Factors, and conduct a small simulation study to better understand the test's operating characteristics. In Section \ref{sec:app}, we illustrate the use of this test with data from a recent study about the absence of the Hawthorne effect. In Section \ref{sec:ANOVA}, we present the analogous non-inferiority test for the $\eta^{2}$ parameter in an ANOVA. We also provide a modified version of this test that allows for the possibility that the variance across groups is unequal.  

\section{A non-inferiority test for the coefficient of determination parameter} \label{sec:R2}

The coefficient of determination, commonly known as $R^{2}$, is a sample statistic used in almost all fields of research. In a linear regression model, the $R^{2}$ is equal to the square of the Pearson correlation coefficient between the observed and predicted outcomes \citep{nagelkerke1991note, zou2003correlation}.   Despite the $R^{2}$ statistic's ubiquitous use, its corresponding population parameter, which we will denote as $P^{2}$,  as in \cite{cramer1987mean}, is rarely discussed.  When considered, it is sometimes known as the ``parent multiple correlation coefficient'' \citep{barten1962note} or the ``population proportion of variance accounted for'' \citep{kelley2007confidence}.  See \cite{cramer1987mean} for a technical discussion. 

%\footnote{Koerts and Abrahamse (1970) \cite{koerts1970correlation} derive $\phi$ as the probability limit of the $R^{2}$ \begin{equation}
 %     \textrm{plim}_{n \to \infty} R^{2}_{n} = \frac{\beta^{'}(m^{-1}X^{'}X)\beta}{\beta^{'}(m^{-1}X^{'}X)\beta + \sigma^{2}} = P^{2} ,
%  \end{equation}

%where $m$ is the XXX, $\beta$ is the XXX, ...}. 

While confidence intervals for $P^{2}$ have been studied by many researchers (e.g., \cite{ohtani2004exact}, \cite{ohtani2000bootstrapping}, \cite{dudgeon2017some}), there has been no consideration (as far as we know) of a non-inferiority test for $P^{2}$.  In this section we will derive such a test and investigate how it compares to a popular Bayesian alternative \citep{rouder2012default}.  Before we continue, let us define some notation.   All technical details are presented in the Appendix.  Let:

\begin{itemize}
\item $N$, be the number of observations in the observed data;
\item $K$, be the number of explanatory variables in the linear regression model;
\item $y_{i}$, be the observed value of random variable $Y$ for the $i$th subject;
\item $x_{ji}$, be the observed value of fixed covariate $X_{j}$, for the $i$th subject, for $k$ in $1, ..., K$; and
\item{$X$, be the $N$ by $K+1$ covariate matrix (with a column of 1s for the intercept; we use the notation $X_{i,\cdot}$ to refer to all $K+1$ values corresponding to the $i$th subject).}
\end{itemize}

\noindent We operate under the standard linear regression assumption that observations in the data are independent and normally distributed with:

\begin{equation} Y_{i} \sim  \; Normal(X_{i,\cdot}^{T}\beta, \sigma^{2}), \quad \quad \forall \; i=1,...,N;
\label{regress}
\end{equation}

\noindent where $\beta$ is a parameter vector of regression coefficients, and $\sigma^{2}$ is the population variance. The parameter  $P^{2}$ represents the proportion of total variance in the population that can be accounted for by knowing the covariates, i.e., by knowing $X$. As such, $P^{2}$ is entirely dependent on the particular design matrix $X$, and we have that: 

\begin{equation} \label{eq:P2}
P^{2} = \frac{\sigma_{XY}^{T}\Sigma^{-1}_{X}\sigma_{XY}}{\sigma^{2}_{Y}},
\end{equation}

\noindent where $\sigma^{2}_{Y}$ is the unconditional variance of $Y$, (note that: $\sigma^{2}_{Y} \ge \sigma^{2}$);  $\sigma_{XY}$ is the vector of population covariances between the $K$ different $X$ variables and $Y$; and $\Sigma_{X}$ is the population covariance matrix of the $K$ different $X$ variables.  The $R^{2}$ statistic estimates the parameter $P^{2}$ from the observed data. See \cite{kelley2007confidence} for a complete derivation of equation (\ref{eq:P2}).

A standard NHST asks whether we can reject the null hypothesis that $P^{2}$ is equal to zero ($H_{0}: P^{2} = 0$).  The $p$-value for this NHST is calculated as: 

\begin{equation} \label{Ftestpval}
p-\textrm{value} = 1 - p_{f}(F; K, N-K-1, 0),
\end{equation}

\noindent where $p_{f}(\cdot \quad ; df_{1}, df_{2}, ncp)$ is the cdf of the non-central $F$-distribution with $df_{1}$ and $df_{2}$ degrees of freedom, and non-centrality parameter $ncp$ (note that $ncp=0$ corresponds to the \emph{central} $F$-distribution); and where:

\begin{equation}
 F = \frac{R^{2}/K}{(1-R^{2})/(N-K-1)}.
 \label{eq:Flm} \end{equation}

\noindent  One can calculate the above $p$-value in \verb|R| with the following code: 

\begin{small}
\noindent \verb|pval = pf(Fstat, df1 = K, df2 = N-K-1, lower.tail = FALSE)|.
\end{small}

A non-inferiority test for $P^{2}$ is asking a different question:  \emph{can we reject the hypothesis that the total proportion of variance in $Y$ attributable to $X$ is greater than or equal to $\Delta$?}  Formally, the hypotheses for the non-inferiority test are:

 $H_{0}: 1 > P^{2} \ge \Delta$,\\
 \indent $H_{1}: 0 \le P^{2} < \Delta$.

\noindent The $p$-value for this non-inferiority test is obtained by inverting the one-sided CI for $P^{2}$ (see Appendix for details), and can be calculated as:

\begin{equation} \label{noninfF}
p-\textrm{value} = p_{f}\left(F; K, N-K-1, \frac{N\Delta}{(1-\Delta)} \right).
\end{equation}

\noindent Note that one can calculate the above $p$-value in \verb|R| with the following code: 

\begin{small}
\noindent \verb|pval = pf(Fstat, df1=K, df2=N-K-1, ncp=(N*Delta)/(1-Delta), lower.tail=TRUE)|.
\end{small}

\textcolor{black}{
Under the assumption that the true value of $P^{2}=0$, for given values of $N$, $K$, and $\Delta$, a simple analytic formula provides a reasonable approximation of the non-inferiority test's statistical power:}

\begin{equation} \label{Rpower2}
power = Pr( \textrm{reject } H_{0}| P^{2}=0) = p_{f}(F^{*}; K, N-K-1, 0),
\end{equation}

\noindent where $F^{*}$ is equal to the $(1-\alpha)$\% critical value of a non-central $F$-distribution with $df_{1}=K$ and $df_{2}=N-K-1$ degrees of freedom, and  non-centrality parameter  $ncp=(N\Delta)/(1-\Delta)$.

\noindent Note that one can calculate the above power estimate in \verb|R| with the following code:

\begin{small}
\noindent \verb|Fstatstar = qf(alpha, df1=K, df2=N-K-1, ncp=(N*Delta)/(1-Delta), lower.tail=TRUE) |
 
\noindent  \verb|power = pf(Fstatstar,df1=K,df2=N-K-1,lower.tail=TRUE)|.
 \end{small}
 
 \noindent It is important to remember that the above tests make two important assumptions about the data:

\begin{itemize}
\item{The data are independent and normally distributed as described in equation (\ref{regress}).}
\item{The values for $X$ in the observed data are fixed and their distribution in the sample is equal (or representative) to their distribution in population of interest. The sampling distribution of $R^{2}$ can be quite different when regressor variables are random; see  \cite{gatsonis1989multiple}.}
\end{itemize}

Ideally, a researcher uses the non-inferiority test to examine a preregistered hypothesis concerning the absence of a meaningful effect.  However, in practice, one might first conduct a NHST (i.e., calculate a $p$-value, $p_{1}$, using equation (\ref{Ftestpval})) and only proceed to conduct the non-inferiority test (i.e., calculate a second $p$-value, $p_{2}$, using equation (\ref{noninfF})) if the NHST fails to reject the null.  Such a two-stage sequential testing scheme has recently been put forward by \cite{campbell2018conditional} under the name of ``conditional equivalence testing'' (CET).  Under the proposed CET scheme, if the first $p$-value, $p_{1}$, is less than the type 1 error $\alpha$-threshold (e.g., if $p_{1} < 0.05$), one concludes with a ``positive'' finding: $P^{2}$ is significantly greater than 0.  On the other hand, if the first $p$-value, $p_{1}$, is greater than $\alpha$ and the second $p$-value, $p_{2}$, is smaller than $\alpha$ (e.g., if $p_{1} \ge 0.05$ and $p_{2} < 0.05$), one concludes with a ``negative'' finding: there is evidence of a statistically significant non-inferiority, i.e., $P^2$ is at most negligible.  If both $p$-values are large, the result is inconclusive: there are insufficient data to support either finding.  

In this paper, we are not advocating for (or against) CET but simply use it to facilitate a comparison with Bayes Factor testing (which also categorizes outcomes as either positive, negative or inconclusive).  Other possible testing strategies available to researchers include: (1) performing only an equivalence test, (2) performing both an equivalence test and a NHST (acknowledging the possibility there is a non-zero, but trivial, effect), and (3) performing a NHST if and only if the equivalence test is not significant. As long as these procedures are chosen and performed transparently (e.g., in a preregistered study) there are scenarios for which all these options can be defended.

\subsection{Comparison to a Bayesian alternative}
\label{sec:bayes}

For linear regression models, based on the work of \cite{liang2008mixtures},  \cite{rouder2012default} propose using Bayes Factors (BFs) to determine whether the data, as summarized by the $R^{2}$ statistic, support the null or the alternative model.  This is a common approach used in psychology studies (e.g., see most recently  \cite{hattenschwiler2019traditional}).  Here we refer to the null model (``Model 0'') and alternative (full) model (``Model 1'') as:

\begin{align}\textrm{Model }\, 0 &: Y_{i} \sim  \quad Normal(\beta_{0}, \sigma^{2}), \quad \quad \forall  i=1,...,N; \\
\textrm{Model }\, 1 &: Y_{i} \sim  \quad Normal(X_{i,\cdot}^{T}\beta, \sigma^{2}), \quad \quad \forall  i=1,...,N;
\label{models}
\end{align}

\noindent where $\beta_{0}$ is the overall mean of $Y$ (i.e., the intercept).

We define the Bayes Factor, $BF_{10}$, as  the probability of the data under the alternative model relative to the probability of the data under the null:

\begin{equation}
BF_{10} = \frac{Pr(Data\,|\,Model\:{1})}{Pr(Data\,|\,Model\:{0})},
\end{equation}

\noindent with the ``10'' subscript indicating that the full model (i.e., ``Model 1'') is being compared to the null model (i.e., ``Model 0'').  The BF can be  easily interpreted.  For example, a $BF_{10}$ equal to 0.10 indicates that the null model is ten times more likely than the full model.  

Bayesian methods require one to define appropriate prior distributions for all model parameters.  \cite{rouder2012default} suggest using ``objective priors'' for linear regression models and explain in detail how one may implement this approach.  We will not discuss the issue of prior specification in detail, and instead point interested readers to \cite{consonni2008compatibility} who provide an in-depth overview of how to specify prior distributions for linear models.

Using the BayesFactor package in \verb|R| \citep{morey2015package} with the function \verb|linearReg.R2stat()|, one can easily obtain a BF corresponding to given values for $R^{2}$, $N$, and $K$.  Since we can also calculate frequentist $p$-values corresponding to given values for $R^{2}$, $N$, and $K$ (see equations  (\ref{Ftestpval}) and (\ref{noninfF})), a comparison between the frequentist and Bayesian approaches is relatively straightforward.  

For three different values of $K$ (=1, 5, 12) and a broad range of values of $N$ (76 values from 30 to 1,000), we calculated the $R^{2}$ values corresponding to a $BF_{10}$ of 1/3 ({``moderate evidence'' in favour of the null model} \citep{jeffreys1961theory}) and of 3 ({``moderate evidence'' in favour of the full model}).  We then proceeded to calculate the corresponding frequentist $p$-values for NHST and non-inferiority testing for the ($R^{2}$, $K$, $N$) combinations. Note that all priors required for calculating the BF were set by simply selecting the default settings of the  \verb|linearReg.R2stat()| function (with  rscale = ``medium''), \textcolor{black}{whereby a noninformative Jeffreys prior is placed on the variance of the normal population, while a scaled Cauchy prior is placed on the standardized effect size; see \citet{morey2015package}.  
}
%For more information on selecting priors for the calculation of BFs, see  \cite{liang2008mixtures} and more recently \cite{quintana2018bayesian}.

\begin{figure}
\includegraphics[width=16cm]{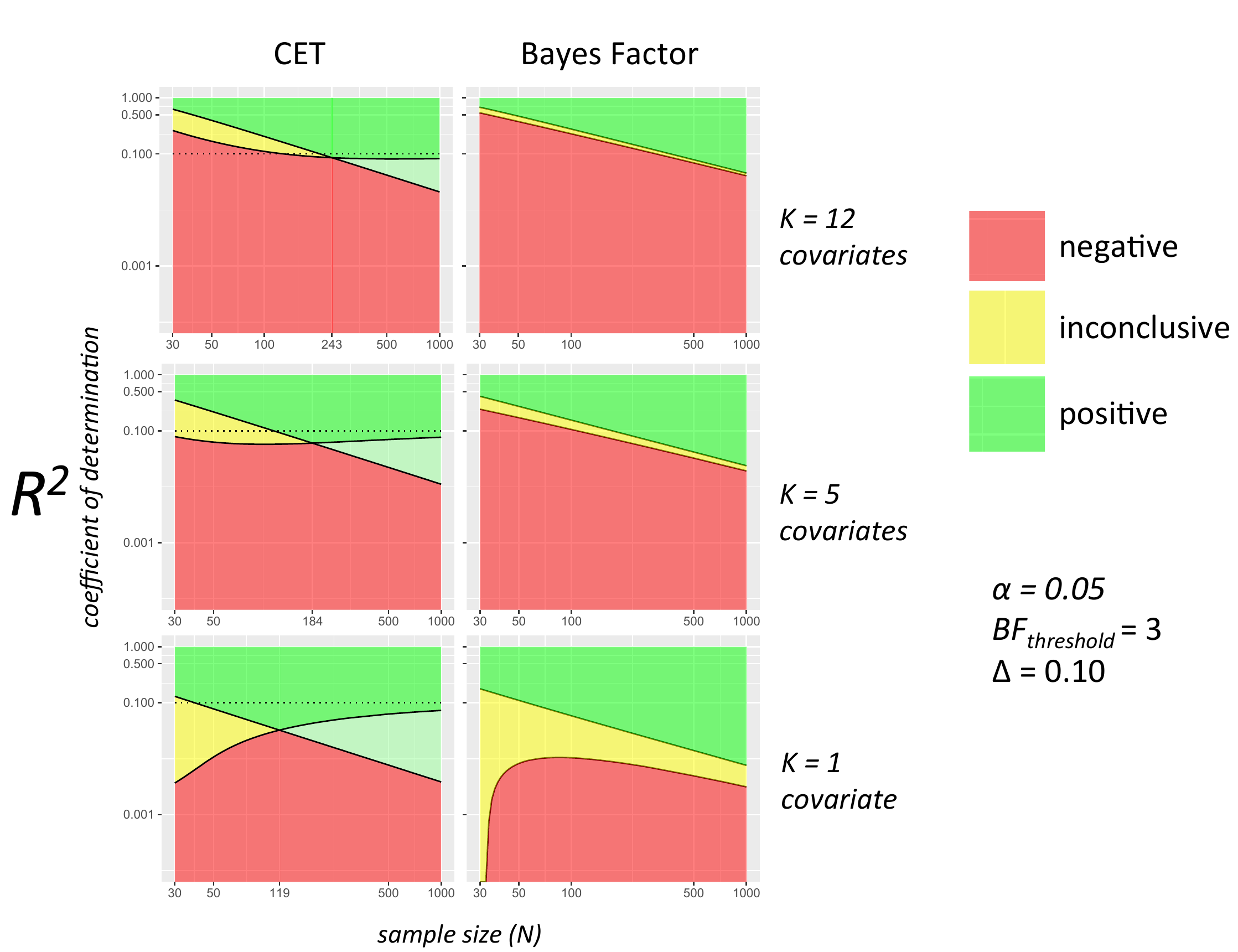}
\caption{Colours indicate the conclusions corresponding to varying levels of $R^{2}$ and $N$ (red=``negative''; yellow = ``inconclusive''; green=``positive''). Left panels shows the frequentist testing scheme with NHST and non-inferiority test ($\Delta=0.10$) and right panels show Bayesian testing scheme with a threshold for the BF of 3.  The significance threshold for frequentist tests is $\alpha=0.05$.  Both vertical-axis ($R^{2}$) and horizontal-axis ($N$) are on logarithmic scales. Note that the ``light-green'' colour corresponds to scenarios for which both the NHST and the non-inferiority $p$-values are less than $\alpha=0.05$.   One could describe the effect in these cases as ``significant yet not meaningful.''  } \label{fig:lakens3}
\end{figure}

The results are plotted in Figure \ref{fig:lakens3}. The left-hand column plots the conclusions reached by frequentist testing (i.e., the CET sequential testing scheme).  For all calculations, we defined $\alpha=0.05$ and $\Delta=0.10$. The right-hand column plots the conclusions reached based on the Bayes Factor with a threshold of 3. 

Each conclusion corresponds to a different colour in the plot: \emph{green} indicates a positive finding (evidence in favour of the full model); \emph{red} indicates a negative finding (evidence in favour of the null model); and \emph{yellow} indicates an inconclusive finding (insufficient evidence to support either model).  Note that we have also included a third colour, light-green.  For the frequentist testing scheme, light-green indicates a scenario where both the NHST $p$-value and the non-inferiority test $p$-value are less than $\alpha=0.05$. The tests reveal that the observed effect size is both statistically significant (i.e., we reject $H_{0}:P^{2}=0$) and statistically smaller than the effect size of interest (i.e., we also reject $H_{0}:P^{2} \ge \Delta$). In these situations, one could conclude that, while $P^2$ is significantly greater than zero, it is likely to be practically insignificant (i.e., a real effect of a negligible magnitude).

 Three observations merit comment:

\noindent (1) When testing with Bayes Factors, there will always exist a combination of values of $R^{2}$ and $N$ that corresponds to an inconclusive result. This is not the case for frequentist testing: the probability of obtaining an inconclusive finding will decrease with increasing $N$, and at a certain point, will be zero. For example, with $K=5$ and any $N>184$, it is impossible to obtain an inconclusive finding regardless of the observed $R^{2}$.

\noindent (2) For $K=1$ covariate, with $N<30$, it is practically impossible to obtain a negative conclusion with the Bayesian approach, and only possible with the frequentist approach (for the equivalence bound of $\Delta=0.10$), if the $R^{2}$ is very very small ($\approx < 0.001$).

\noindent (3) For $K=12$ covariates, with $N<50$, the frequentist testing scheme obtains a negative conclusion in situations when $R^{2}>\Delta$.  This may seem rather odd but can be explained by the fact that $R^{2}$ is ``seriously biased upward in small samples'' \citep{cramer1987mean}.

%Are the assumptions needed for CET the same as those needed for BF?  What about random regressors?

%Research on $R^{2}$ for Bayesian models continues, see Gelman et al. (2018) \cite{gelman2018r}.

\textcolor{black}{Based on this comparison of BFs and frequentist tests, we can speculate that, given the same data, both approaches will often provide one with the same overall conclusion.  In Section \ref{sec:simstudy2}, we investigate this further by means of a simulation study.}

\subsection{Simulation study 1}
\label{sec:simstudy1}

\begin{figure}[p]
    \centering
        \includegraphics[width=14cm]{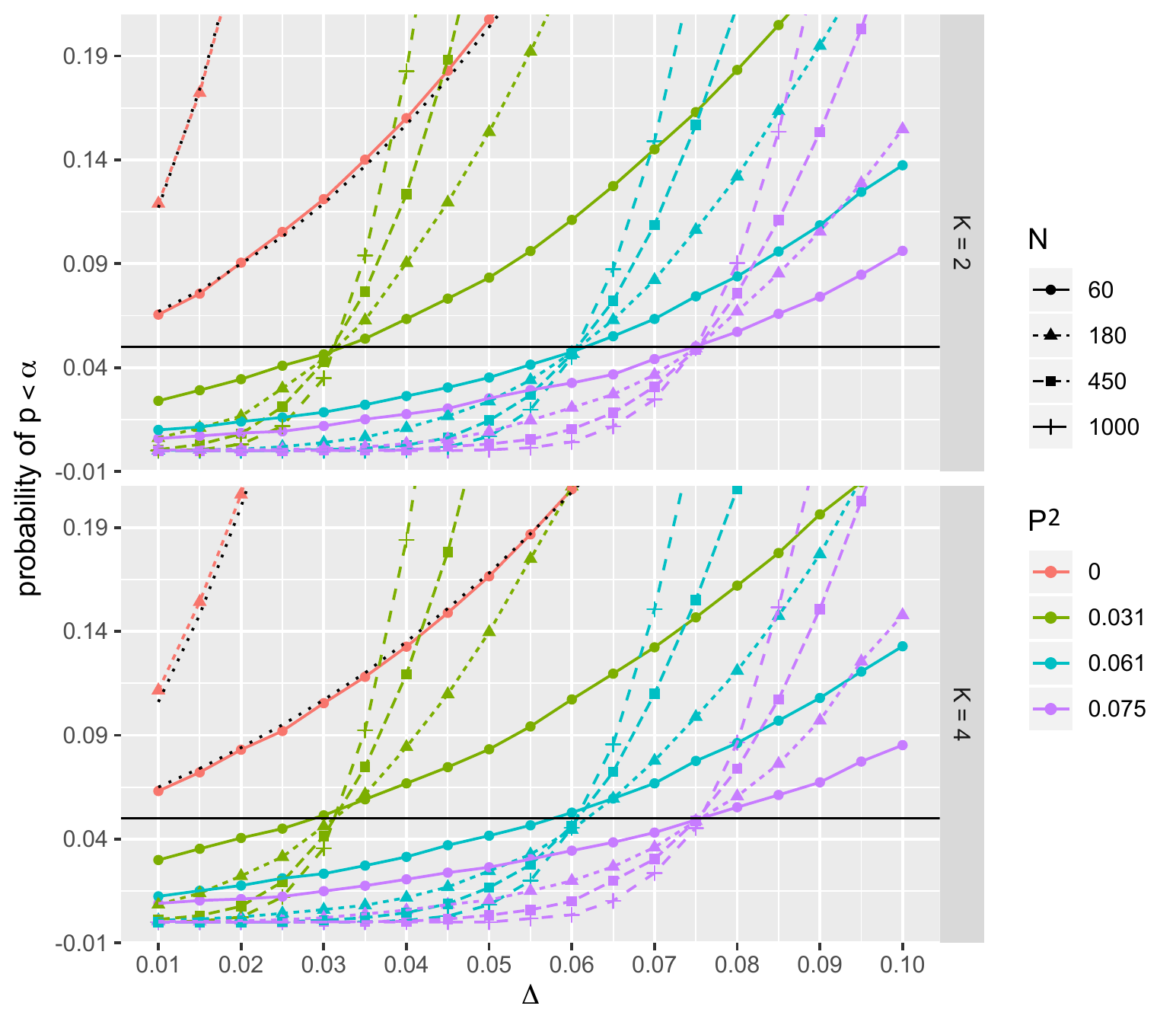}
    \caption{Simulation study results.  Upper panel shows results for $K=2$; lower panel shows results for $K=4$.  Both plots are presented with a restricted vertical-axis to better show the type 1 error rates.  The solid horizontal black line indicates the desired type 1 error of $\alpha=0.05$.   The dotted black curves plot numbers calculated using equation (\ref{Rpower2}) for estimating power.  For each of thirty-two configurations, we simulated 50,000 unique datasets and calculated a non-inferiority $p$-value with each of 19 different values of $\Delta$ (ranging from 0.01 to 0.10).}
    \label{fig:type1}
\end{figure}

 \begin{figure}[p]
    \centering
    \includegraphics[width=14cm]{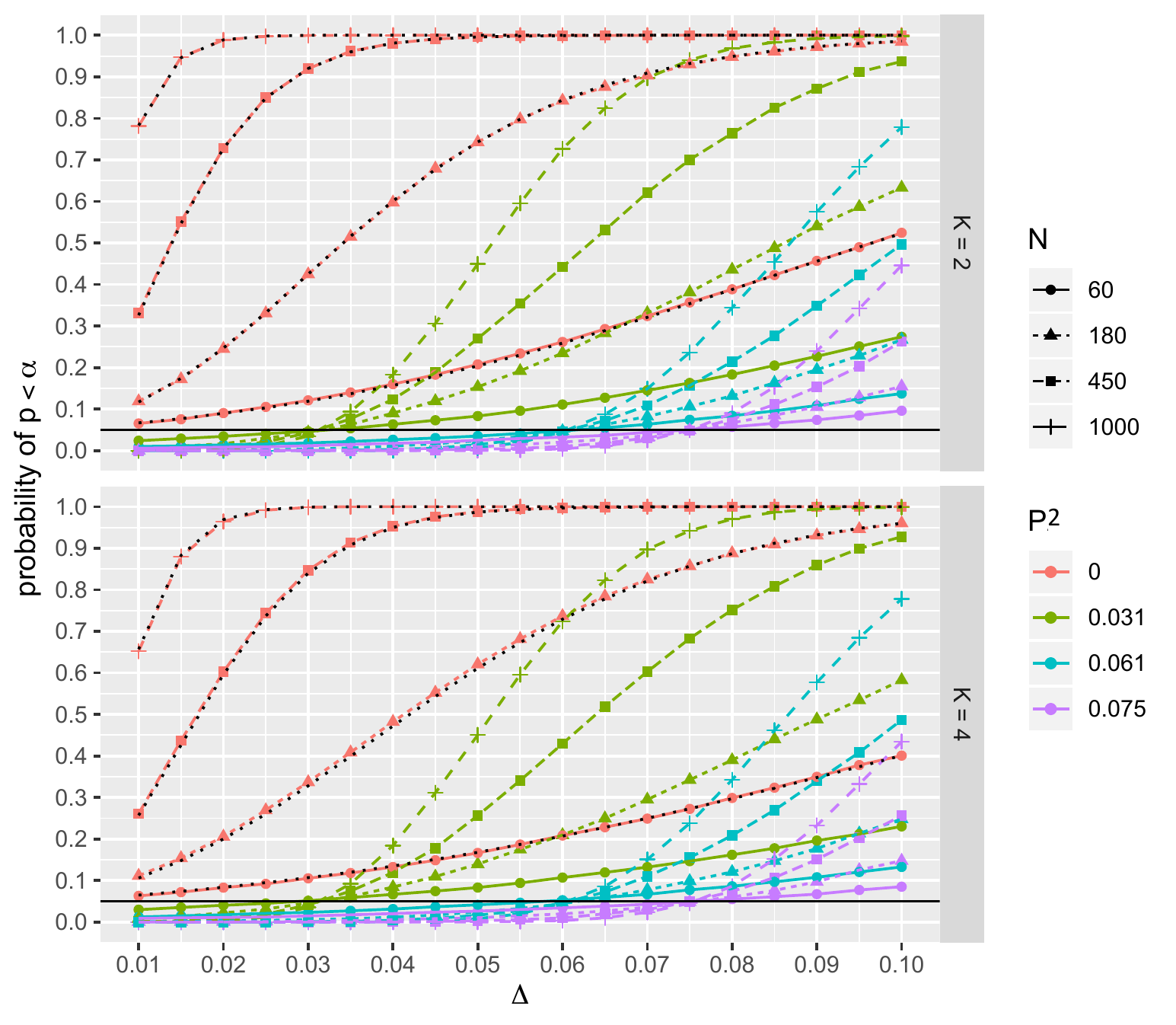}
    \caption{Simulation study, complete results.  Upper panel shows results for $K=2$; Lower panel shows results for $K=4$.  The solid horizontal black line indicates the desired type 1 error of $\alpha=0.05$.  The dotted black curves plot numbers calculated using equation (\ref{Rpower2}) for estimating power.  For each of thirty-two configurations, we simulated 50,000 unique datasets and calculated a non-inferiority $p$-value with each of 19 different values of $\Delta$ (ranging from 0.01 to 0.10).}
    \label{fig:power}
\end{figure}

We conducted a simple simulation study in order to better understand the operating characteristics of the non-inferiority test and to confirm that the test has correct type 1 error rates.  We simulated data for each of \textcolor{black}{twenty-four} scenarios, one for each combination of the following parameters:
 
\begin{itemize}
    \item one of four sample sizes: $N=60$, $N=180$, $N=540$, or, $N=1,000$;
    \item one of two designs with $K=2$, or $K=4$ binary covariates, (with an orthogonal, balanced design), and with $\beta=(0.0, 0.2, 0.3)$ or $\beta=(0.0, 0.2, 0.2, -0.1, -0.2)$; and 
    \item one of three variances: $\sigma^{2}=0.4$ ,$\sigma^{2}=0.5$, or $\sigma^{2}=1.0$.
\end{itemize}

\noindent Depending on the particular values of $K$ and $\sigma^{2}$, the true coefficient of \textcolor{black}{determination} for these data is either $P^{2}=0.031$, $P^{2}=0.061$, or $P^{2}=0.075$.  Parameters for the simulation study were chosen so that we would consider a wide range of values for the \textcolor{black}{ sample size (representative of those sample sizes commonly used in the psychology literature; see \cite{kuhberger2014publication}, \cite{fraley2014n}, and \cite{marszalek2011sample}) and so as to obtain three unique values for $P^{2}$ approximately evenly spaced between 0 and 0.10.}  

\textcolor{black}{We also simulated data from eight additional scenarios where $P^{2}=0$.  This will allow us to confirm that the proposed function (equation (\ref{Rpower2})) for estimating power is accurate.  These additional scenarios were based on the following:}

\begin{itemize}
    \item \textcolor{black}{one of four sample sizes: $N=60$, $N=180$, $N=540$, or, $N=1,000$;}
    \item \textcolor{black}{one of two designs with $K=2$, or $K=4$ binary covariates, with $\beta=(0.0, 0.0, 0.0)$ or $\beta=(0.0, 0.0, 0.0, 0.0, 0.0)$; and $\sigma^{2}=1.0.$}
\end{itemize}

\textcolor{black}{ For each of the thirty-two configurations, we simulated 50,000 unique datasets and calculated a non-inferiority $p$-value with each of 19 different values of $\Delta$ (ranging from 0.01 to 0.10).  We then calculated the proportion of these $p$-values less than $\alpha=0.05$.  We specifically chose to conduct 50,000 simulation runs so as to keep computing time within a reasonable limit while also reducing the amount of Monte Carlo standard error to a negligible amount (for looking at type 1 error with $\alpha=0.05$, Monte Carlo SE will be approximately $0.001 \approx \sqrt{0.05(1-0.05)/50,000}$); see \cite{morris2019using}.}

Figure \ref{fig:type1} plots the results with a restricted vertical axis to better show the type 1 error rates. Figure \ref{fig:power} plots the results against the unrestricted vertical \textcolor{black}{ axis.  Both plots also show dotted black curves which correspond to the numbers obtained using equation (\ref{Rpower2}) for calculating power.}

We see that when the equivalence bound $\Delta$ equals the true effect size (i.e., 0.031, 0.061, or 0.075), the type 1 error rate is exactly 0.05, as it should be, for all $N$. This situation represents the boundary of the null hypothesis, i.e. $H_{0}: \Delta \le P^{2}$.
As the equivalence bound increases beyond the true effect size (i.e., $\Delta > P^{2}$), the alternative hypothesis is then true and it becomes possible to correctly conclude equivalence. 

\textcolor{black}{
As expected, the power of the test increases with larger values of $\Delta$, larger values of $N$, and smaller values of $K$.  Also, in order for the test to have substantial power, the $P^{2}$ must be substantially smaller than $\Delta$.  The agreement between the red curves ($P^{2}=0$) and the dotted black curves suggests that the analytic function presented in equation (\ref{Rpower2}) provides a fairly accurate approximation of the statistical power.}

\subsection{Simulation study 2}
\label{sec:simstudy2}

We conducted a second simulation study to compare the operating characteristics of testing with the JZS-BF relative to testing with the frequentist CET approach.  (Note that the frequentist and Bayesian testing schemes we consider are but two of many options available to researchers.  For example, one could consider a Bayesian approach that uses an interval-based null; see \cite{kruschke2018bayesian}.)

In this simulation study, frequentist conclusions were based on setting $\Delta$ equal to either 0.01, or 0.05, or 0.10; and with $\alpha$=0.05.  JZS-BF conclusions were based on an evidence threshold of either 3, 6, or 10.  A threshold of 3 can be considered ``moderate evidence,'' a threshold of 6 can be considered ``strong evidence,'' and a threshold of 10 can be considered ``very strong evidence'' \citep{jeffreys1961theory, wagenmakers2011psychologists}.  Note that for the simulation study here we examine only the ``fixed-$n$ design'' for BF testing; see \cite{schonbrodt2016bayes} for details.  Also note that, as in Section \ref{sec:bayes}, all priors required for calculating the BF were set by simply selecting the default settings of the  \verb|linearReg.R2stat()| function (with  rscale = ``medium''), \textcolor{black}{whereby a noninformative Jeffreys prior is placed on the variance of the normal population, while a scaled Cauchy prior is placed on the standardized effect size; see \citet{morey2015package}.  
}

We simulated datasets for \textcolor{black}{64} unique scenarios.  We considered the following parameters:
 
\begin{itemize}
    \item one of sixteen sample sizes: $N=20$,   $N=30$,   $N=42$,   $N=60$,   $N=88$,   $N=126$,   $N=182$,   $N=264$,   $N=380$,   $N=550$,   $N=794$,   $N=1,148$,   $N=1,658$,   $N=2,396$,   $N=3,460$, or   $N=5,000$;
    \item one of two designs with $K=4$ binary covariates (with an orthogonal, balanced design), with either $\beta=(0.0, 0.2, 0.2, -0.1, -0.2)$ or $\beta=(0.0, 0.0, 0.0, 0.0, 0.0)$;  
    \item one of three variances: $\sigma^{2}=9.0$$, \sigma^{2}=1.0$, or $\sigma^{2}=0.5$.
\end{itemize}

\noindent Note that for the $\beta=(0.0, 0.0, 0.0, 0.0, 0.0)$ design, we only consider one value for $\sigma^{2}=1.0$.  Depending on the particular design and $\sigma^{2}$, the true coefficient of \textcolor{black}{determination} for these data is either $P^{2}=0.000$, $P^{2}=0.004$, $P^{2}=0.031$, or $P^{2}=0.061$.

\begin{figure}[p]
    \centering
    \includegraphics[width=15cm]{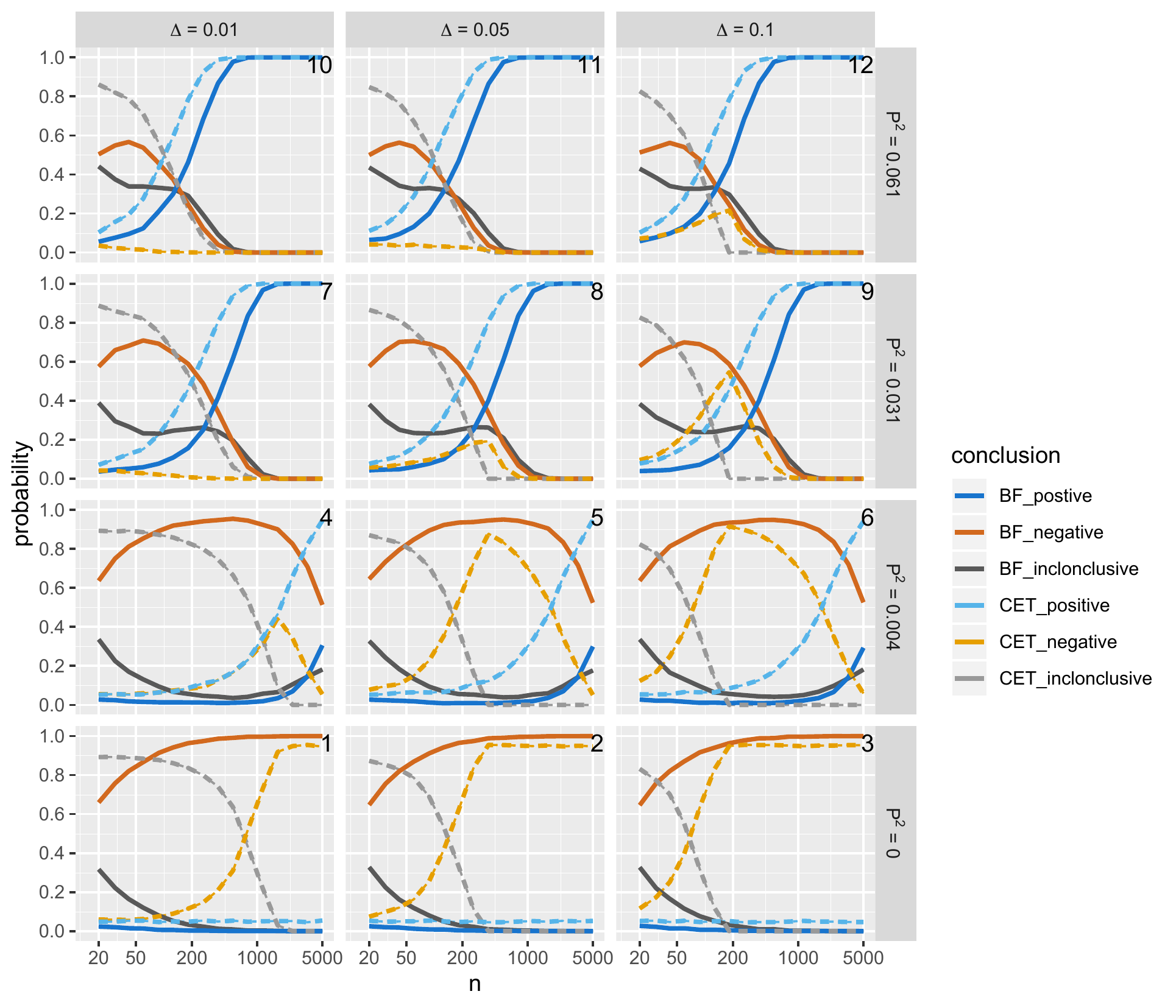}
    \caption{\textbf{Simulation study 2, complete results for BF threshold of 3.}  The probability of obtaining each conclusion by Bayesian testing scheme (JZS-BF with fixed sample size design,  BF threshold of 3:1) and CET ($\alpha=0.05$).  Each panel displays the results of simulations with for different values of $\Delta$ and $P^{2}$.  Note that all solid lines and the dashed blue line do not change for different values of $\Delta$.}
    \label{fig:ss2_BF3}
\end{figure}

\begin{figure}[p]
    \centering
    \includegraphics[width=15cm]{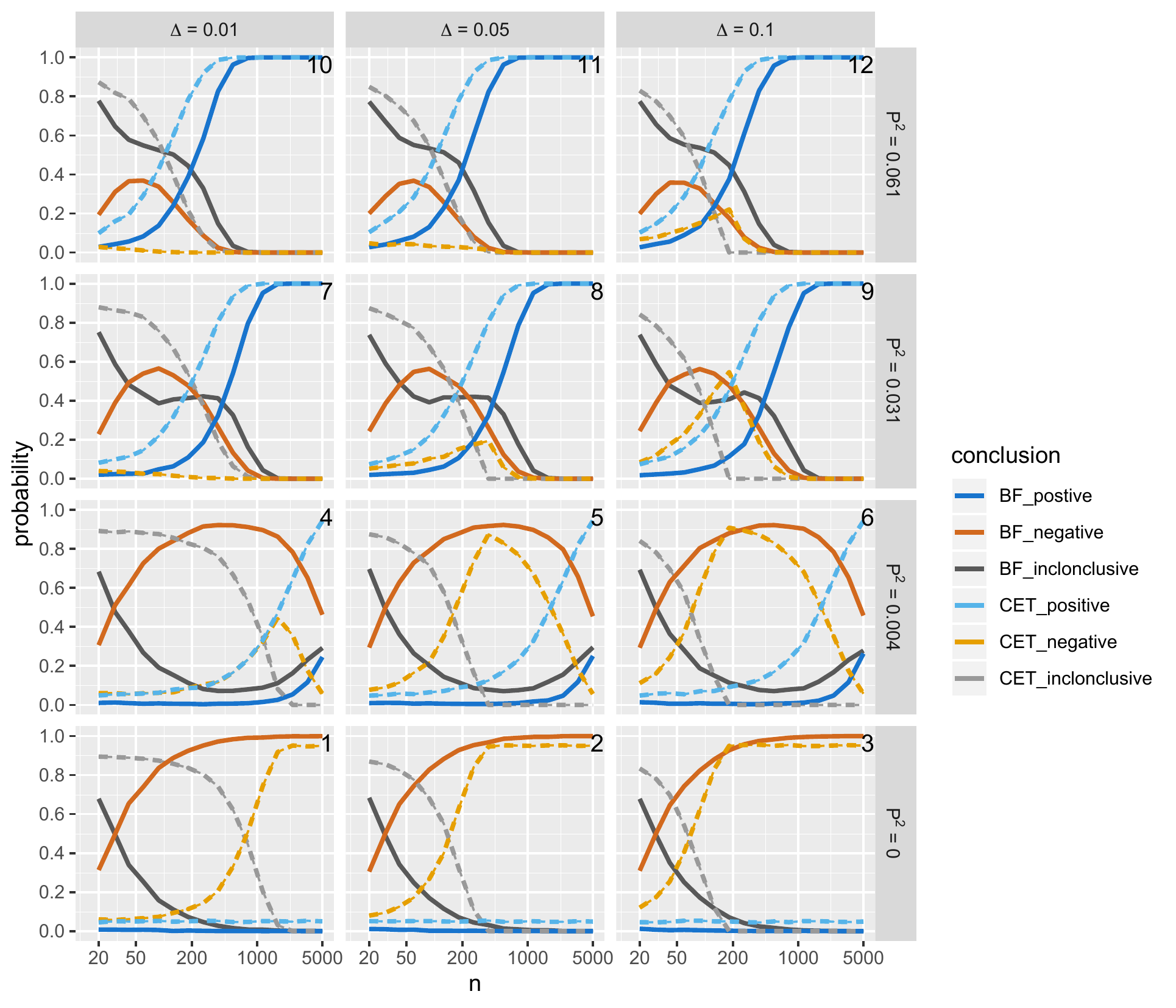}
    \caption{\textbf{Simulation study 2, complete results for BF threshold of 6.}  The probability of obtaining each conclusion by Bayesian testing scheme (JZS-BF with fixed sample size design,  BF threshold of 6:1) and CET ($\alpha=0.05$).  Each panel displays the results of simulations with for different values of $\Delta$ and $P^{2}$.  Note that all solid lines and the dashed blue line do not change for different values of $\Delta$.}
    \label{fig:ss2_BF6}
\end{figure}

\begin{figure}[p]
    \centering
    \includegraphics[width=15cm]{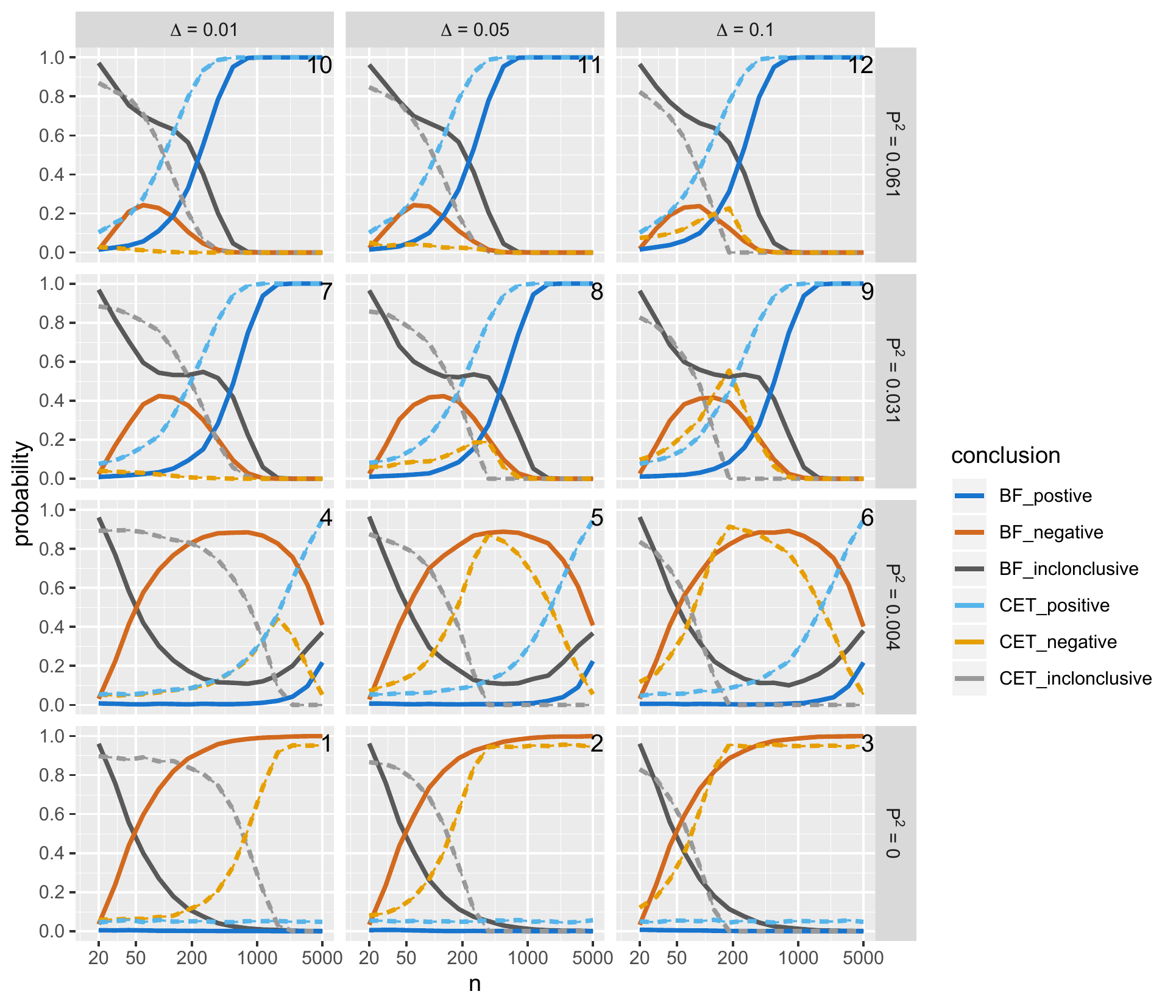}
    \caption{\textbf{Simulation study 2, complete results for BF threshold of 10.}  The probability of obtaining each conclusion by Bayesian testing scheme (JZS-BF with fixed sample size design,  BF threshold of 10:1) and CET ($\alpha=0.05$).  Each panel displays the results of simulations with for different values of $\Delta$ and $P^{2}$.  Note that all solid lines and the dashed blue line do not change for different values of $\Delta$.}
    \label{fig:ss2_BF10}
\end{figure}

 For each simulated dataset, we obtained frequentist $p$-values, JZS-BFs and declared the result to be positive, negative or inconclusive accordingly.   Results are presented in Figures \ref{fig:ss2_BF3}, \ref{fig:ss2_BF6} and \ref{fig:ss2_BF10} and are based on 5,000 distinct simulated datasets per scenario.  We are also interested in how often the two approaches will reach the same overall conclusion: \emph{averaging over all 64 scenarios, how often on average will the Bayesian and frequentist approaches reach the same conclusion given the same data?}  Table \ref{tab:agree} displays the the average rate of agreement between the Bayesian and frequentist methods. 

Three observations merit comment:

 \begin{itemize}
 \item{With an evidence  threshold of 3 or of 6, the JZS-BF requires substantially less data to reach a negative conclusion than the frequentist scheme in most cases.  However, with an evidence threshold of 10 and $\Delta=0.10$, both methods are approximately equally likely to deliver a negative conclusion.  Note that, the probability of reaching a negative result with CET will never exceed 0.95 since the NHST is performed first and will reach a positive result with probability $1-\alpha$; see dashed orange lines in Figures \ref{fig:ss2_BF3}, \ref{fig:ss2_BF6}, and \ref{fig:ss2_BF10} - panels 1, 2, and 3.}
 
  \item{While the JZS-BF requires less data to reach a conclusive result when the sample size is small (see how the solid black curve drops more rapidly than the dashed grey line), there are scenarios in which larger sample sizes will surprisingly reduce the likelihood of the BF obtaining a conclusive result (see how the solid black curve drops abruptly then rises slightly as $n$ increases for $P^{2}=$ 0.004, and 0.031; see for example, Figure \ref{fig:ss2_BF10} - panels 4 and 7). }
  
   \item{The JZS-BF is always less likely to deliver a positive conclusion (see how the dashed blue curve is always higher than the solid blue curve).  In scenarios like the ones we considered, the JZS-BF may require larger sample sizes for reaching a positive conclusion and thus may be considered ``less powerful'' in a traditional frequentist sense.}
   \end{itemize}

Based on our comparison of BFs and frequentist tests, we can confirm that, given the same data, both approaches will often provide one with the same overall conclusion.  The level of agreement however is highly sensitive to the choice of $\Delta$ and the choice of the BF evidence threshold, see Table \ref{tab:agree}.  The highest level of agreement, recorded at 0.80, is when $\Delta=0.10$ and the BF evidence threshold is equal to 10.  In contrast, when $\Delta=0.01$ and the BF evidence threshold is 3, the two approaches will only deliver the same conclusion half of the time.  Table \ref{tab:disagree} shows that the two approaches rarely arrive at entirely contradictory conclusions.  In less than 6\% of cases, did we observe one approach arrive at a positive conclusion while the other approach arrived at a negative conclusion when faced with the same exact data.

The results of this second simulation study suggest that, depending on how they are configured, the JZS-BF and CET may operate very similarly.  Think of JZS-BF and CET as two pragmatically similar, yet philosophically different, tools for making ``trichotomous significance-testing decisions.''  This simulation study result is reassuring since it suggests that the conclusions obtained from frequentist and Bayesian testing will rarely lead to substantial disagreements.

% latex table generated in R 3.5.3 by xtable 1.8-4 package
% Wed Nov  6 13:54:52 2019
\vspace{1cm}
\begin{table}[ht]
\centering
\begin{tabular}{|r|ccc|}
  \hline
 & BF thres. = 3 & BF thres. = 6 & BF thres. = 10 \\ 
  \hline
$\Delta = 0.10$ & 0.719& 0.767& 0.800 \\ 
$\Delta = 0.05$ & 0.628& 0.683& 0.735 \\ 
$\Delta = 0.01$ & 0.485& 0.538& 0.594 \\ 
   \hline
\end{tabular}
\caption{ Averaging over all 96 scenarios and over all 5,000 Monte Carlo simulations per scenario, how often on average did the Bayesian and frequentist approaches reach the same conclusion?  Numbers in the above table represent the average proportion of simulated datasets (averaged over $64\times5,000 = 320,000$ unique datasets) for which the Bayesian and frequentist methods arrive at the same conclusion.}
\label{tab:agree}
\end{table}

% latex table generated in R 3.5.3 by xtable 1.8-4 package
% Wed Nov  6 13:54:52 2019
\vspace{1cm}
\begin{table}[ht]
\centering
\begin{tabular}{|r|ccc|}
  \hline
 & BF thres. = 3 & BF thres. = 6 & BF thres. = 10 \\ 
  \hline
$\Delta = 0.10$ & 0.055 & 0.042 & 0.034 \\ 
$\Delta = 0.05$ & 0.055 & 0.042 & 0.034 \\ 
$\Delta = 0.01$ & 0.056 & 0.042 & 0.035 \\ 
   \hline
\end{tabular}
\caption{ Averaging over all 96 scenarios and over all 5,000 Monte Carlo simulations per scenario, how often on average did the Bayesian and frequentist approaches strongly disagree in their conclusion?  Numbers in the above table represent the average proportion of simulated datasets (averaged over $64\times5,000 = 320,000$ unique datasets) for which the Bayesian and frequentist methods arrived at completely opposite (one positive and one negative) conclusions.}
\label{tab:disagree}
\end{table}

\section{Practical Example: Evidence for the absence of a Hawthorne effect}\label{sec:app}

 \cite{mccambridge2019randomized} tested the hypothesis that participants who know that the behavioral focus of a study is alcohol related will modify their consumption of alcohol while under study.  The phenomenon of subjects modifying their behaviour simply because they are being observed is commonly known as the Hawthorne effect \citep{stand2000hawthorne}.  

The researchers conducted a three-arm individually randomized trial online among students in four New Zealand universities.  The three groups were: group A (control), who were told they were completing a lifestyle survey; group B, who were told the focus of the survey was alcohol consumption; and group C, who additionally answered 20 questions on their alcohol use and its consequences before answering the same lifestyle questions as Groups A and B.  The prespecified primary outcome was a subject's self-reported volume of alcohol consumption in the previous 4 weeks (units = number of standard drinks).  This measure was recorded at baseline and after one month at follow-up.  See Table \ref{tab:haw} for a summary  of the data from McCambridge at el. (2019).

% latex table generated in R 3.3.3 by xtable 1.8-2 package
% Tue Apr 23 15:01:36 2019
\begin{table}[ht]
\centering
\begin{tabular}{lrrrr}
  \hline
& & \textbf{baseline} & \textbf{followup} & \textbf{difference} \\ 
  \hline
\textbf{A}& $N$ & 1795 & 1483 & 1483 \\ 
&  mean & 24.60 & 18.39 & -5.13 \\ 
&  sd & 31.80 & 23.32 & 24.56 \\ 
\hline
\textbf{B}&   $N$ & 1852 & 1532 & 1532 \\ 
&  mean & 23.83 & 17.48 & -5.64 \\ 
&  sd & 31.79 & 23.81 & 21.77 \\ 
\hline
\textbf{C}&   $N$ & 1825 & 1565 & 1565 \\ 
&  mean & 23.03 & 17.45 & -4.79 \\ 
&  sd & 30.65 & 23.21 & 25.17 \\ 
\hline
\textbf{Total} &   $N$ & 5472 & 4582 & 4580 \\ 
&  mean & 23.82 & 17.77 & -5.19 \\ 
&  sd & 31.42 & 23.44 & 23.88 \\ 
   \hline
\end{tabular}
\caption{Summary of the data from McCambridge at el. (2019).  The table summarizes the prespecified primary outcome, a subject's self-reported volume of alcohol consumption in the previous 4 weeks (units = number of standard drinks).  This measure was recorded at baseline and after one month at follow-up in each of the three experimental groups.}
\label{tab:haw}
\end{table}

The data were analyzed by \cite{mccambridge2019randomized} using a linear regression model with repeated measures fit by generalized estimating equations (GEE) and an ``independence'' correlation structure.  For a NHST of the overall experimental group effect, the researchers obtained a $p$-value of 0.66.  Based on this result, \cite{mccambridge2019randomized} conclude that ``the groups were not found to change differently over time.''  

We note that this linear regression model fit by GEE is just one of many potential models one could use to analyze this data; see \citet{yang2001efficiency}.  Three (among many) other reasonable alternative approaches include (1) a linear model using only the follow-up responses (without adjustment for the baseline measurement); (2) a linear model using the follow-up responses as outcome with a covariate adjustment for the baseline measurement; and (3) a linear model using the difference between follow-up and baseline responses as outcome.  These three approaches yield $p$-values of 0.45, 0.56, and 0.61, respectively.  None of these $p$-values suggest rejecting the null hypothesis.  In order to show evidence ``in favour of the null,'' we turn to our proposed non-inferiority test.

We fit the data ($N=4,580$) with a linear regression model using the difference between follow-up and baseline responses as the outcome, and the group membership as a categorical covariate, $K=2$.  We then consider the non-inferiority test for the coefficient of determination parameter (see Section \ref{sec:R2}), with  $\Delta =0.01$.  This test asks the following question: does the overall experimental group effect account for less than 1\% of the variability explained in the outcome? 

The choice of $\Delta =0.01$ represents our belief that any Hawthorne effect explaining less than 1\% of the variability in the data would be considered negligible.  For reference, Cohen (1988) describes a $R^{2}=0.0196$ as ``a modest enough amount, just barely escaping triviality'';  and more recently,  \cite{fritz2012effect} consider associations explaining ``1\% of the variability'' as  ``trivial.''  It is up to researchers to provide a justification of the equivalence bound before they collect the data. Researchers can specify the non-inferiority margin based on theoretical predictions (for example derived from a computational model), based on a cost-benefit analysis, or based on discussions among experts who decide which effects are too small to be considered meaningful. 

%In general, subject matter expertise will play a role in determining appropriate values for $\Detla$, e.g. a systematic review of the literature on studies of Hawthorne effect  \cite{mccambridge2014systematic} could be useful in this case.

  We obtain a $R^{2}=0.000216$ and can calculate the $F$-statistic with equation (\ref{eq:Flm}):

\begin{align}
F &=& \frac{R^{2}/K}{(1-R^{2})/(N-K-1)} \\
&=& \frac{0.000216/2}{(1-0.000216)/(4580-2-1)} \\
&=& \frac{0.000108}{0.000218}\\
&=& 0.49
\end{align}
 
To obtain a $p$-value for the non-inferiority test, we use equation (\ref{noninfF}):

\begin{align}
p-\textrm{value} &=&  p_{f}\left(F; K, N-K-1, \frac{N\Delta}{(1-\Delta)} \right)\\
&=& p_{f}\left(0.49; 2, 4580-2-1, \frac{4580 \cdot 0.01}{(1-0.01)} \right) \\
&=& 1.13 \times 10^{-9}
\end{align}

This result, $p$-value $= 1.13 \times 10^{-9}$, suggests that we can confidently reject the null hypothesis that $P^{2}>0.01$.  We therefore conclude that the data are most compatible with no important effect.  For comparison, the Bayesian testing scheme we considered in Section \ref{sec:bayes} obtains a Bayes Factor of $B_{10}=0.00284 = 1/352$.  The \verb|R| code for these calculations is presented in the Appendix.

%(Recently some researchers have called for a ``multiverse approach'' \cite{steegen2016increasing}... )

%GRE testing data:

%\verb|http://www.ets.org/s/gre/pdf/snapshot.pdf|

\section{A non-inferiority test for the ANOVA $\eta^{2}$ parameter}
 \label{sec:ANOVA}
 
 Despite being entirely equivalent to linear regression \citep{gelman2005analysis}, the fixed effects (or ``between subjects") analysis of variance (ANOVA) continues to be the most common statistical procedure to test the equality of multiple independent population means in many fields \citep{plonsky2017multiple}.  The non-inferiority test considered earlier in the linear regression context will now be described in an ANOVA context for evaluating the equivalence of multiple independent groups.  \textcolor{black}{We must emphasize that the two versions are essentially the same test described with different names.}  Note that all tests developed and discussed in this paper are only for \emph{between}-subject ANOVA designs and cannot be applied to \emph{within}-subject designs. Extensions to within and mixed designs is a fruitful effort for future research.

 Equivalence/non-inferiority tests for comparing group means in an ANOVA have been proposed before.  For example,  \cite{rusticus2011applying} list several examples of studies that used ANOVA to compare multiple groups in which non-significant findings are incorrectly used to conclude that groups are comparable.  The authors emphasize the problem (``a statistically non-significant finding only indicates that there is not enough evidence to support that two (or more) groups are statistically different'') and offer an equivalence testing solution based on CIs.  Unfortunately, conclusions of equivalence are based only on CIs which the authors warn may be ``too wide'' \citep{rusticus2011applying}.  
 
 In another proposal,  \cite{wellek2010testing} considered simultaneous equivalence testing for several parameters to test group means.  However, this strategy may not necessarily be more efficient than the rather inefficient strategy of multiple pairwise comparisons; see the conclusions of  \cite{pallmann2017simultaneous}.    Koh and Cribbie (2013) (see also \cite{cribbie2009tests}) consider two different omnibus tests.  These are presented as non-inferiority tests for $\varphi^{2}$, a parameter closely related to the population signal-to-noise parameter, $s2n$; (note that $s2n = \varphi^{2}/N$, where $N$ is the total sample size). Unfortunately, the use of these tests is limited by the fact that the population parameters $\varphi^{2}$ and $s2n$ are not commonly used in analyses since their units of measurement are rather arbitrary.
 
 In this section, we consider a non-inferiority test for the population effect-size parameter, $\eta^{2}$, a standardized effect size that is commonly used in the social sciences \citep{kelley2007confidence}. The parameter $\eta^{2}$ represents the proportion of total variance in the population that can be accounted for by knowing the group level.  The use of commonly used standardized effect sizes is recommended in order to facilitate future meta-analysis and the interpretation of results \citep{lakens2013calculating}.  Note that $\eta^{2}$ is analogous to the $P^{2}$ parameter considered earlier in the linear regression context in Section \ref{sec:R2}.  Also note that the non-inferiority test we propose is entirely equivalent to the test for $\varphi^{2}$ proposed by  \cite{koh2013robust}.  It is simply a re-formulation of the test in terms of the $\eta^{2}$ parameter.
 
 Before going forward, let us define some basic notation.  All technical details are presented in the Appendix. Let $Y$ represent the continuous (normally distributed) outcome variable, and $X$ represent a fixed categorical variable (i.e., group membership).  Let $N$ be the total number of observations in the observed data, $J$ be the number of groups (i.e., factor levels in $X$), and $n_{j}$ be the number of observations in the $j$th group, for $j$ in 1,..., $J$.  We will consider two separate cases, one in which the variance within each group is equal, and one in which variance is heterogeneous.

 Typically, one will conduct a standard $F$-test to determine whether one can reject the null hypothesis that $\eta^{2}$ is equal to zero ($H_{0}: \eta^{2} = 0$).  The $p$-value is calculated as:

\begin{equation}
p-\textrm{value} = 1 - p_{f}(F; J-1, N-J, 0),
\label{nhst_anova}
\end{equation}

\noindent where, as in Section \ref{sec:R2}, $p_{f}(\cdot \quad ; df_{1}, df_{2}, ncp)$ is the cdf of the non-central F-distribution with $df_{1}$ and $df_{2}$ degrees of freedom, and non-centrality parameter, $ncp$ (note that $ncp=0$ corresponds to the \emph{central} $F$-distribution); and where:

\begin{equation}
 F = \frac{\sum_{j=1}^{J}n_{j}(\bar{y}_{j} - \bar{y})^{2}/(J-1) }{\sum_{j=1}^{J}\sum_{i=1}^{n_{j}} (y_{ij} - \bar{y}_{j})^{2}/(N-J)}.
 \end{equation}

\noindent One can calculate the above $p$-value using \verb|R| with the following code:

\begin{small}
\noindent \verb|pval = pf(Fstat,df1=J-1,df2=N-J,lower.tail=FALSE)|.
\end{small}

A non-inferiority test for $\eta^{2}$ asks a different question:  \emph{can we reject the hypothesis that the total amount of variance in $Y$ attributable to group membership is greater than $\Delta$?} Formally, the hypotheses for the non-inferiority test are written as:

 $H_{0}: 1 > \eta^{2} \ge \Delta$,\\
 \indent $H_{1}: 0 < \eta^{2} \le \Delta$.

\noindent If we reject $H_{0}$, we reject the hypothesis that there are meaningful differences between the group means ($\mu_{j}$, $j=1,...,J$), in favour of the hypothesis that the group means are considered practically equivalent. The $p$-value for this test is obtained by inverting the one-sided CI for $\eta^{2}$ (see Appendix for details) and can be calculated as:

\begin{equation}
p-\textrm{value} = p_{f}\left(F; J-1, N-J, \frac{N\Delta}{(1-\Delta)} \right).
\label{hom_p}
\end{equation}

\noindent Note that one can calculate the above $p$-value using  \verb|R| with the following code:

\begin{small}
\noindent \verb|pval = pf(Fstat,df1=J-1,df2=N-J,ncp=N*Delta/(1-Delta),lower.tail=TRUE)|.
\end{small}

Under the assumption that the true value of $\eta^{2}=0$, for given values of $N$, $J$, and $\Delta$, a simple analytic formula provides an estimate for the non-inferiority test's statistical power:

\begin{equation} \label{Rpower}
power = Pr( \textrm{reject } H_{0}| \eta^{2}=0) = p_{f}(F^{*}; J-1, N-J, 0),
\end{equation}

\noindent where $F^{*}$ is equal to the $(1-\alpha)$\% critical value of a non-central $F$-distribution with $df_{1}=J-1$ and $df_{2}=N-J$ degrees of freedom, and  non-centrality parameter  $ncp=(N\Delta)/(1-\Delta)$.

\noindent Note that one can calculate the above power estimate in \verb|R| with the following code:
\begin{small}
 \verb|Fstatstar = qf(alpha, df1 = J-1, df2 = N-J, ncp = (N*Delta)/(1-Delta), lower.tail = TRUE) |

\noindent \verb|power = pf(Fstatstar, df1 = J-1, df2 = N-J, lower.tail = TRUE)|.
\end{small} 

  The non-inferiority test for $\eta^{2}$  makes the following three important assumptions about the data:

\begin{itemize}
\item{The outcome data  are independent and normally distributed.}
\item{The proportions of observations for each group (i.e., $n_{j}/N$, for $j=1,...J)$ that are in the observed data are equal to the proportions that are in the total population of interest.}
\item{The variance within each group is equal (homogeneous variance).}
\end{itemize}

\subsection{A non-inferiority test for ANOVA with heterogeneous variance}

With regards to the third assumption above, we can modify the above non-inferiority test in order to allow for the possibility that the variance is unequal across groups (heterogeneous variance).  Welch's $F$-test statistic is calculated as (see Appendix for details; see also \cite{delacre2018taking}):

	  \begin{equation}
	F^{'} = \frac{\sum_{j}^{J}\hat{w}_{j}(y_{j}-\bar{y}^{'})^2/(J-1)}{1 + \frac{2(J-2)}{J^{2}-1} \sum_{j=1}^{J}((n_{j}-1)^{-1})(1- \frac{\hat{w}_{j}}{\hat{W}})},
	\end{equation}

\noindent where $\hat{w}_{j} = {n_{j}}/{s_{j}^{2}}$, with $s_{j}^{2} = \sum_{i=1}^{n_{j}} ((y_{ij}-\bar{y}_{j})^2)/(n_{j}-1)$, for $j=1,...,J$; and where  $\hat{W}=\sum_{j=1}^{J}\hat{w}_{j}$, and $\bar{y}^{'} = \sum_{j=1}^{J}(\hat{w}_{j}\bar{y}_{j})/ \hat{W}$, for $j=1,...,J$.

Then, the $p$-value for a non-inferiority test ($H_{0}: 1 > \eta^{2} \ge \Delta$) in the case of heterogeneous variance is:

\begin{equation}
p-\textrm{value} = p_{f}(F^{'}; J-1, df^{'}, \frac{N\Delta}{(1-\Delta)} ).
\label{het_p}
\end{equation}	

\noindent where:

\begin{equation}
    df^{'} =\frac{J^{2}-1}{3\sum_{j=1}^{J}((n_{j}-1)^{-1})(1- {\hat{w}_{j}}/{\hat{W}})^{2}}.
\end{equation}

\noindent The above $p$-value can be calculated using \verb|R| with the following code: 

\begin{small}
 \begin{verbatim}
aov1 <- oneway.test(y ~ x, var.equal = FALSE)        
Fprime <- aov1$statistic
dfprime <-  aov1$parameter[2]
pval = pf(Fprime, J-1, df2 = dfprime, ncp = (Delta*N)/(1-Delta), lower.tail=TRUE)
 \end{verbatim}
 \end{small}

 For the heterogeneous case the population effect size parameter, $\eta^{2}$, is defined slightly differently than for the homogeneous case (see Appendix for details).   Based on the simulation studies of \cite{koh2013robust}, we can recommend that the non-inferiority test based on the Welch's $F^{'}$ statistic (i.e., the test with $p$-value calculated from equation (\ref{het_p})) is almost always preferable (with regards to the statistical power and type 1 error rate) to the test which requires an assumption of homogeneous variance (i.e., the test with $p$-value calculated from equation (\ref{hom_p})).  \textcolor{black}{ This agrees with similar recommendations for using Welch's $t$-test (e.g., \cite{delacre2017psychologists, ruxton2006unequal}).  We also point interested readers to the related work of \cite{jan2019extended}. }

\textcolor{black}{ Some might advocate for a  two-step procedure: using the homoscedastic version as a default and only moving to the Welch version as needed based on a preliminary test for homogeneity of variance.  However, problems with this kind of preliminary testing (e.g., first testing for equality of variances, then deciding upon which test to use) have been identified (e.g., \cite{zimmerman2004conditional, zimmerman2004note, campbell2014consequences}), and as such, the use of the two-step procedure cannot be recommended.}

%\citet{rochon2012test} investigate the consequences of conducting a preliminary test for normality (e.g., the Shapiro-Wilk test).  The authors conclude that while ``[f]rom a formal perspective, preliminary testing for normality is incorrect and should therefore be avoided,'' in practice, ``preliminary testing does not seem to cause much harm, at least for the cases we have investigated.''

\subsection{The absence of a Hawthorne effect (ANOVA)}

For the absence of a Hawthorne effect example we considered earlier in Section \ref{sec:app}, note that we can easily analyze the data in an ANOVA framework (and obtain the identical result).  The standard ANOVA output is summarized in Table \ref{tab:anova}.

% latex table generated in R 3.5.3 by xtable 1.8-4 package
% Wed Nov  6 10:39:54 2019
\begin{table}[h!]
\centering
\begin{tabular}{lrrrr}
  \hline
 & df & Sum Sq & Mean Sq & $F$ value \\ 
  \hline
Experimental Group & 2 & 562.77 & 281.38 & 0.49  \\ 
  Residuals                  & 4577 & 2609820.50 & 570.20 &  \\ 
   \hline
   &&&& $\hat{\eta}^{2} = 0.000216$\\
\end{tabular}
\caption{ANOVA summary of the absence of a Hawthorne effect example data.}
\label{tab:anova}
\end{table}

\noindent In this case, with $\Delta=0.01$, a non-inferiority test can be conducted for $\eta^{2}$ ($H_{0}: 1 > \eta^{2} \ge \Delta$) and a $p$-value is calculated using equation (\ref{hom_p}) as follows:

\begin{align}
p-\textrm{value} &=& p_{f}\left(F; J-1, N-J, \frac{N\Delta}{(1-\Delta)} \right) \\
&=& p_{f}\left(0.49; 3-1, 4580-3, \frac{4580\cdot0.01}{(1-0.01)} \right) \\
&=& 1.13 \times 10^{-9}.
\label{example_aov}
\end{align}

\section{Conclusion}\label{sec:conclsn}

In this paper we presented a statistical method for non-inferiority testing of standardized omnibus effects commonly used in linear regression and ANOVA.  We also considered how frequentist non-inferiority testing, and equivalence testing more generally, offer an attractive alternative to Bayesian methods for ``testing the null.''  We recommend that all researchers specify an appropriate non-inferiority margin and, at a minimum, plan to use the proposed non-inferiority tests in the event that a standard NHST fails to reject the null.  In cases when the sample size is very large, the non-inferiority test can be useful to detect effects that are statistically significant but not meaningful.  

\textcolor{black}{We wish to emphasize that the use of equivalence/non-inferiority tests should not rule out the complementary use of confidence intervals. Indeed,
confidence intervals can be extremely useful for highlighting the stability (or lack of stability) of a given estimator, whether that be the $R^{2}$, $\hat{\eta}^{2}$ or any other statistic.  Perhaps one advantage of equivalence/non-inferiority testing over confidence intervals may be that testing can improve the interpretation of null results \citep{parkhurst2001statistical, hauck1986proposal}.  By clearly distinguishing between what is a ``negative'' versus an ``inconclusive'' result, equivalence testing serves to simplify the long ``series of searching questions''  necessary to evaluate a ``failed outcome'' \citep{pocock2016primary}.  In our opinion, the best interpretation of data will be when using both tools together and our proposal simply serves to ``extend the arsenal of confirmatory methods rooted in the frequentist paradigm of inference'' \citep{wellek2017critical}.}

Note that our current non-inferiority test for $P^{2}$ in a standard multivariable linear regression is limited to comparing the ``full model'' to the ``null model.''  As such, the test is not suitable for comparing two nested models.  For example, we cannot use the test to compare a ``smaller model'' with only the baseline measure as a covariate, with a ``larger model'' that includes both baseline measure \emph{and} group membership as covariates.  Equivalence testing for comparing two nested models will be addressed in future work in which we will consider a non-inferiority test for the increase in $R^{2}$ between a smaller model and a larger model.  Related work includes that of \citet{algina2007confidence,algina2008note}.  We also wish to further investigate non-inferiority testing for ANOVA with \emph{within}-subject designs, following the work of \cite{rose2018new}.  It would also be interesting (and worthwhile) to develop non-inferiority tests to tackle the $R^{2}$ calculated for generalized linear mixed-effects models \citep{nakagawa2013general}.

 There is a great risk of bias in the scientific literature if researchers only rely on statistical tools that can reject null hypotheses, but do not have access to statistical tools that allow them to reject the presence of meaningful effects.  Most recently, Amrhein et al. (2019) express great concern with the the practice of statistically non-significant results being ``interpreted as indicating `no difference' or `no effect' '' \citep{amrhein2019scientists}; see also \citet{altman1995statistics}. Equivalence tests provide one approach to improve current research practices by allowing researchers to falsify their predictions concerning the presence of an effect.

\textcolor{black}{By specifying equivalence bounds, researchers can design studies that yield informative answers both when the alternative hypothesis is true, and when the null hypothesis is true. Using equivalence tests to reject the presence of meaningful effects makes it possible to conclude predictions are falsified, and thus might be a way to mitigate problems that are caused by publication bias. However, equivalence tests can also be abused. Researchers might be tempted to specify the equivalence bounds after looking at the data such that the equivalence test is guaranteed to be statistically significant. Ideally, equivalence bounds are pre-specified and documented in a preregistration document that is made available when a manuscript is submitted for publication to avoid flexibility in the data analysis.  Equivalence bounds should always be justified and independent of the observed data.  Weak justifications weaken the statistical inference.} Thinking about what would falsify your prediction is a crucial step when designing a study, and specifying a smallest effect size of interest and performing an equivalence test provides one way to answer that question. 
\vskip 0.52in

%\paragraph*{Available Code -}  All the code used in this paper and relevant materials are made available in an OSF repository (https://osf.io/3q2vh/), DOI 10.17605/OSF.IO/3Q2VH.  

%\paragraph*{Acknowledgements}

%Thank you to Prof. Paul Gustafson for the helpful advice with preliminary drafts.  Thank you to Prof. John Petkau for the generous help with editing.

 \section{Appendix}

 \subsection{Linear Regression: further details and R-code.}

The $R^{2}$ statistic estimates the parameter $P^{2}$ from the observed data:

\begin{equation} 
 R^{2} = 1 - \frac{SS_{RES}}{SS_{TOT}},
\end{equation}

\noindent where $SS_{RES}=
\sum_{i=1}^{N}\left(y_{i}-\hat{y}_{i}\right)^{2}$
, and $SS_{TOT}=
\sum_{i=1}^{N}\left(y_{i}-\overline{y}\right)^{2}
$; with $\hat{y} = 
X^{T}\left(X^{\prime} X\right)^{-1} X^{\prime} Y
$, and $\bar{y} = \sum_{i=1}^{N}y_{i}/N$.

The R-code for analysis of the  \cite{mccambridge2019randomized} data is:

\begin{verbatim}
    
Xmatrix <- model.matrix(totaldrinking.diff  ~ group, data= side_data)
lmmodel <- lm(totaldrinking.diff  ~ group , data= side_data)

R2 <- summary(lmmodel)$r.squared
Fstat <- summary(lmmodel)$fstatistic[1]
K <- dim(Xmatrix)[2] - 1
N <- dim(Xmatrix)[1]
Delta <- 0.01
 
pf(Fstat,df1=K,df2=N-K-1,ncp=(N*Delta)/(1-Delta),lower.tail=TRUE)

linearReg.R2stat(N=N, p=K, R2= R2, simple=TRUE)


\end{verbatim}

The code below replicates the results published in McCambridge et al. (2019). Note that there appears to be a typo in the published table whereby the $p$-values 0.89 and 0.86 are switched.

\begin{verbatim}


Hdata$group<-relevel(Hdata$group,"A")

mod0 <- geeglm(totaldrinking ~ + group+t, 
id= participant_ID, corstr="independence", data= Hdata, x=TRUE)
mod1 <- geeglm(totaldrinking ~ group*t + group+t, 
id= participant_ID, corstr="independence", data= Hdata)
(anova(mod1,mod0))
summary(mod1)$coefficients

Hdata$group<-relevel(Hdata$group,"C")
mod1a <- geeglm(totaldrinking ~ group*t + group+t, 
id= participant_ID, corstr="independence", data= Hdata)
summary(mod1a)

\end{verbatim}

 \subsection{ANOVA with homogeneous variance: further details.}
 
The true population group mean for group $j$ is denoted  $\mu_{j}$, for $j$ in 1,..., $J$; and we denote the group effects as  $\tau_{j} = \mu_{j} - \mu$, where $\mu$ is the overall weighted population mean, $\mu = (\sum_{j=1}^{J}\mu_{j}n_{j})/N$.  These parameters are estimated from the observed data by the corresponding sample group means: $\hat{\mu}_{j} = \bar{y}_{j} = (\sum_{i=1}^{n_{j}}{y_{i}})/n_{j}$, for $j$ in 1,...,$J$; and the overall sample mean: $\hat{\mu} = \bar{y}=(\sum_{j=1}^{J}{\bar{y}_{j}}n_{j})/N$.  

We operate under the assumption that the data is normally distributed such that:
\begin{equation} Y_{i,j} \sim \; Normal(\mu_{j}, \sigma^{2}_w),\; \forall \; j=1,...,J, \quad \forall i=1,...,n_{j},
\end{equation}

\noindent where $\sigma_{w}^{2}$ denotes the variance within groups. 
We also define the variance between groups as $\sigma_{b}^{2} = \sum_{j=1}^{J}n_{j}(\mu_{j} - \mu)^{2}/N$. Finally, the total population variance is defined as $\sigma_{t}^{2} = \sigma_{b}^{2} + \sigma_{w}^{2}$. The corresponding sums of squares are estimated from the data: $SS_{b} = \sum_{j=1}^{J}n_{j}(\bar{y}_{j} - \bar{y})^{2}$; $SS_{w} = \sum_{j=1}^{J}\sum_{i=1}^{n_{j}} (y_{ij} - \bar{y}_{j})^{2}$; and $SS_{t}= SS_{b} + SS_{w}$. 

\noindent Recall that the ANOVA F-test statistic is calculated as:

\begin{equation}
 F = \frac{SSb/df_{b} }{SS_{w}/df_{w}} = \frac{\sum_{j=1}^{J}n_{j}(\bar{y}_{j} - \bar{y})^{2}/(J-1) }{\sum_{j=1}^{J}\sum_{i=1}^{n_{j}} (y_{ij} - \bar{y}_{j})^{2}/(N-J)},
 \end{equation}
 
 \noindent where $df_{b} = J-1$, and $df_{w}= N-J$. The $F$ statistic follows an F distribution with degrees of freedom $df_{b}$ for the numerator, and $df_{w}$ degrees of freedom for the denominator.   
 %Typically, a $p$-value is calculated from the $F$-test as: $p$-value =  $1-p_{f}(F, df_{b},  df_{w})$.
 
 The population effect size, $\eta^{2} \in [0,1]$, is a parameter that represents the amount of variance in the outcome variable, $Y$, that is explained by the group membership, (i.e., knowing the level of the factor $X$), and is defined as:
 
 %(analogous to the $\phi$ in the regression context)
 
 \begin{equation}
 \eta^{2} = \frac{\sigma_{b}^{2}}{\sigma_{t}^{2}} = \frac{\sigma_{b}^{2}}{\sigma_{b}^{2} + \sigma_{w}^{2}} =  1 - \frac{\sigma_{w}^{2}}{\sigma_{t}^{2}}
 \end{equation}

  \noindent  We can estimate the population parameter $\eta^{2}$ from the observed data using the sample statistic,  $\hat{\eta}^{2}$, as follows: $\hat{\eta}^{2} =  SS_{b}/SS_{t}$. It is well known that $\hat{\eta}^{2}$ is a biased estimate for  ${\eta}^{2}$.  However, alternative estimates (including $\hat{\epsilon}^{2} = (SS_{b}-dfb \cdot MS_{w})/SS_{t}$, and  $\hat{\omega}^{2} = (SS_{b}-df_{b}\cdot MS_{w})/(SS_{t}+MS_{w})$) are also biased; see \cite{okada2013omega} for more details (note that there is a typo in eq. 5 of \cite{okada2013omega}). 
  %While all three of these estimators ($ \hat{\eta}^{2} $, $\hat{\epsilon}^{2}$, and $\hat{\omega}^{2}$) are biased, asymptotically (i.e., for very large $n$) they are all unbiased, see Maxell (1981).

The population effect size parameter $\eta^{2}$ is closely related to the signal-to-noise ratio parameter, $s2n= \sigma^{2}_{b}/\sigma^{2}_{w}$, and to the non-centrality parameter, $\Lambda = \sum_{j=1}^{J}n_{j}\tau_{j}^{2}/\sigma^{2}_{w} =  {N\sigma_{b}^{2}}/{\sigma_{w}^{2}}$.  Consider the following equality:

\begin{equation}
    \eta^{2} = \frac{s2n}{1-s2n} = \frac{\Lambda}{\Lambda + N}.
\end{equation}

 The non-centrality parameter, $\Lambda$, is estimated from the data as: $\hat{\Lambda} = (n-1)SS_{b}/SS_{w}$, and we can easily calculate a one-sided $(1-\alpha)\%$ confidence interval (CI), $[0, \Lambda_{U}]$, by ``pivoting'' the cumulative distribution function (cdf); see \cite{kelley2007confidence} Section 2.2 and references therein.  This requires solving (numerically) the following equation for $\Lambda_{U}$:

\begin{equation}
p_{f}(F; df_{1}=df_{b}, df_{2}=df_{w}, ncp= \Lambda_{U}) = \alpha,
\end{equation}

\noindent where $p_{f}(\cdot \quad ; df_{1}, df_{2}, ncp)$ is the cdf of the non-central F-distribution with $df_{1}$ and $df_{2}$ degrees of freedom, and non-centrality parameter, $ncp$. The values for $F$, $df_{b}$, $df_{w}$, are calculated from the data as defined above. The solution, $\Lambda_{U}$, will be the upper confidence bound of $\Lambda$, such that: $Pr(\Lambda<\Lambda_{U}) = \alpha$.

 As detailed in \cite{kelley2007confidence} (note that there is a typo in eq. 55 of \cite{kelley2007confidence}: $\Lambda_{L}$ in the numerator should be $\Lambda_{U}$) one can convert the bounds of the CI for $\Lambda$ into bounds for a CI for $\eta^{2}$.  The upper limit of a one-sided CI for $\eta^{2}$ is: $\eta^{2}_{U} = \Lambda_{U}/(\Lambda_{U}+N)$. As such, we have that  $Pr(\eta^{2} \le \frac{\Lambda_{U}}{\Lambda_{U} + N}) = 1 - \alpha$.

\subsection{ANOVA with heterogeneous variance: further details.}

As above, the true population group mean for group $j$  is denoted  $\mu_{j}$, for $j$ in 1,...,$J$.  We now define:

\begin{equation} Y_{i,j} \; \sim \; \; Normal(\mu_{j}, \sigma^{2}_{w,j}), \; \forall j=1,...,J, \quad \forall i=1,...,n_{j},
\end{equation}

\noindent  and define $w_{j} = n_{j}/\sigma^{2}_{w,j}$, and $W=\sum_{j=1}^{J}w_{j}$, and finally $\bar{\mu}^{'} = \sum_{j=1}^{J}(w_{j}\mu_{j})/W$. 

Recall that a Welch F-test statistic is calculated as:

	  \begin{equation}
	F^{'} = \frac{\sum_{j}^{J}\hat{w}_{j}(\bar{y}_{j}-\bar{y}^{'})^2/(J-1)}{1 + \frac{2(J-2)}{J^{2}-1} \sum_{j=1}^{J}((n_{j}-1)^{-1})(1- {\hat{w}_{j}}/{\hat{W}})^{2}},
	\end{equation}

\noindent where $\hat{w}_{j} = {n_{j}}/{s_{j}^{2}}$, with $s_{j}^{2} = \sum_{i=1}^{n_{j}} ((y_{ij}-\bar{y}_{j})^2)/(n_{j}-1)$, for $j=1,...,J$; and where  $\hat{W}=\sum_{j=1}^{J}\hat{w}_{j}$, and $\bar{y}^{'} = \sum_{j=1}^{J}(\hat{w}_{j}\bar{y}_{j})/ \hat{W}$, for $j=1,...,J$.  
 
\cite{levy1978some} proposed an approximate non-null distribution for the $F^{'}$ statistic such that $F^{'}$ follows a non-central $F$-distribution with $df_{1}=J-1$ and $df_{2}=df^{'}$ degrees of freedom, and non-centrality parameter, $\Lambda^{'} = \sum_{j=1}^{J}w_{j}(\mu_j-\bar{\mu}^{'})^2$; see also \cite{jan2014sample}. The degrees of freedom for this case are defined as: $df_{1}=J-1$, and:

\begin{equation}
    df^{'} =\frac{J^{2}-1}{3\sum_{j=1}^{J}((n_{j}-1)^{-1})(1- {\hat{w}_{j}}/{\hat{W}})^{2}}
\end{equation}
 
 We will therefore define our population effect size parameter for the heterogeneous case as:
	
\begin{equation}
    \eta^{2'} = \frac{\Lambda^{'}}{\Lambda^{'} + N}.
\end{equation}

Note that in the case of homogeneous variance (i.e., when $\sigma^{2}_{w,j}=\sigma^{2}_{w,k}, \forall j,k$ in $1,...,J$), we have $\Lambda^{'}=\Lambda$ and $\eta^{2'}=\eta^{2}$. The $p$-value for the non-inferiority test ($H_{0}: \eta^{2'} > \Delta$) in the case of heterogeneous variance is:

\begin{equation}
p-\textrm{value} = p_{f}\left(F^{'}; J-1, df^{'}, ncp = \frac{N\Delta}{(1-\Delta)} \right).
\end{equation}

\vspace{20cm}

\bibliography{truthinscience} 
%Where the bibliography will be 

\end{document}